\documentclass[preprint,pre,aps,amsfonts,amsmath,showpacs,superscriptaddress,nofootinbib, draft]{revtex4}
\usepackage{bm}
\begin{document}
\title{Note on the Transition to Intermittency for the Exponential of the Square of a Steinhaus Series}
\author{Philippe Mounaix}
\email{mounaix@cpht.polytechnique.fr}
\author{Pierre Collet}
\email{collet@cpht.polytechnique.fr}
\affiliation{Centre de Physique Th\'eorique, UMR 7644 du CNRS, Ecole
Polytechnique, 91128 Palaiseau Cedex, France.}
\date{\today}
\begin{abstract}
Intermittency of $\mathcal{E}_N(x,g)=\exp\lbrack g\vert S_N(x)\vert^2\rbrack$ as $N\rightarrow +\infty$ is investigated on a $d$-dimensional torus $\Lambda$, when $S_N(x)$ is a finite Steinhaus series of $(2N+1)^d$ terms normalized to $\langle\vert S_N(x)\vert^2\rangle =1$. Assuming ergodicity of $\mathcal{E}_N(x,g)$ as $N\rightarrow +\infty$ in the domain $g<1$, where $\lim_{N\rightarrow +\infty}\langle\mathcal{E}_N(g)\rangle$ exists, transition to intermittency is proved as $g$ increases past the threshold $g_{th}=1$. This transition goes together with a transition from (assumed) ergodicity at $g<g_{th}$ to  a regime where $\lim_{N\rightarrow +\infty}\lbrack\vert\Lambda\vert\langle\mathcal{E}_N(g)\rangle\rbrack^{-1}\int_{\Lambda}\mathcal{E}_N(x,g)\, d^dx=0$ at $g>g_{th}$. In this asymptotic sense one can say that ergodicity is lost as $g$ increases past the value $g=1$.
\end{abstract}
\maketitle
%
\newtheorem{lemma}{Lemma}
\newtheorem{proposition}{Proposition}
\section{Introduction}\label{sec1}
This paper is the first of a series devoted to studying intermittency of the solution to the random PDE,
\begin{equation}\label{eq1.1}
\left\lbrace
\begin{array}{l}
\partial_t\mathcal{E}_N(x,t)-\frac{i}{2m}\Delta\mathcal{E}_N(x,t)=
\lambda\vert S_N(x,t)\vert^2\mathcal{E}_N(x,t), \\
t\ge 0,\ x\in\Lambda\subset\mathbb{R}^d,\ {\rm and}\ \mathcal{E}_N(x,0)=1,
\end{array}
\right.
\end{equation}
as $N\rightarrow +\infty$, where $S_N(x,t)$ is a sum of $(2N+1)^d$ modes with i.i.d. random phases. Here $\lambda >0$ is the coupling constant and $m\ne 0$ is a complex mass with ${\rm Im}(m)\ge 0$.

For ${\rm Im}(m)=0$ and ${\rm Re}(m)\ne 0$,\ (\ref{eq1.1}) models the scattering of an incoherent laser by an optically active medium. In this context, a simpler version in which $S_N(x,t)$ is approximated by a Gaussian random field was first considered by Akhmanov {\it et al.} in nonlinear optics\ \cite{ADP}, and by Rose and DuBois in laser-plasma interaction\ \cite{RD1}. The latter investigated the divergence of the average solution to\ (\ref{eq1.1}) heuristically and numerically. The same problem was analyzed from a more rigorous mathematical point of view in\ \cite{ADLM},\ \cite{ML}, and\ \cite{MCL}. Going beyond the Gaussian approximation requires further specification of $S_N(x,t)$. In realistic models laser light is represented by a superposition of a finite number of monochromatic beamlets with i.i.d. random phases\ \cite{RD2}. The class of $S_N(x,t)$ considered in\ (\ref{eq1.1}) is a straighforward generalization of those models. For every $N<+\infty$, $S_N(x,t)$ is bounded and according to\ \cite{MCL} there is no divergence of the average of $\mathcal{E}_N(x,t)$. To get interesting results from\ \cite{MCL} one needs to work out the $N\rightarrow +\infty$ limit of their theory. A possible alternative approach is suggested in the introduction of\ \cite{ADLM}. It is explained there that the divergence of the average solution to\ (\ref{eq1.1}) indicates a change in the nature of $\mathcal{E}_N(x,t)$ which undergoes a transition to intermittency. Taking it the other way round leads to characterize $\mathcal{E}_N(x,t)$ by its intermittency, the divergence of its average taking a back seat. That is the approach followed in this work. Note that the problem\ (\ref{eq1.1}) is complementary to the one considered in\ \cite{MCL} in which $N$ is fixed and $S_N(x,t)$ is a sum of $N$ independent Gaussian random variables. By the central limit theorem, the results of\ \cite{MCL} are expected to coincide with those of the present work in the limit $N\rightarrow +\infty$.

For a given $t>0$, $\mathcal{E}_N(x,t)$ is said to be intermittent if, for every integer $p\ge 1$, the space average of $\vert\mathcal{E}_N(x,t)\vert^p$ is almost surely determined by higher and higher and more and more widely spaced peaks of $\vert\mathcal{E}_N(x,t)\vert$ as $N\rightarrow +\infty$. Intermittency of $\mathcal{E}_N(x,t)$ can be inferred from the almost sure chain of strong inequalities\ \cite{Mol},
\begin{equation}\label{eq1.2}
1\ll\frac{1}{\vert\Lambda\vert}\int_{\Lambda}\vert\mathcal{E}_N(x,t)\vert\, d^dx
\ll\cdots\ll\left\lbrack\frac{1}{\vert\Lambda\vert}\int_{\Lambda}\vert\mathcal{E}_N(x,t)\vert^p d^dx\right\rbrack^{1/p}\ll\cdots
\end{equation}
where $f(N)\ll g(N)$ means $\liminf_{N\rightarrow +\infty}g(N)/f(N)=+\infty$. To prove that\ (\ref{eq1.2}) does provide a sufficient condition for intermittency, choose for every realization for which\ (\ref{eq1.2}) is fulfilled a sequence $\lbrace f_p(N)\rbrace$ such that
\begin{equation}\label{eq1.3a}
1\ll f_0(N)\ll\frac{1}{\vert\Lambda\vert}\int_{\Lambda}\vert\mathcal{E}_N(x,t)\vert\, d^dx,
\end{equation}
and, for every integer $p\ge 1$,
\begin{equation}\label{eq1.3b}
\left\lbrack\frac{1}{\vert\Lambda\vert}
\int_{\Lambda}\vert\mathcal{E}_N(x,t)\vert^p d^dx\right\rbrack^{1/p}\ll f_p(N)\ll \left\lbrack\frac{1}{\vert\Lambda\vert}
\int_{\Lambda}\vert\mathcal{E}_N(x,t)\vert^{p+1} d^dx\right\rbrack^{1/(p+1)}.
\end{equation}
from\ (\ref{eq1.3a}) and\ (\ref{eq1.3b}) it follows that, $\forall p\ge 1$,
$$
\frac{1}{\vert\Lambda\vert}\int_{\Lambda}\vert\mathcal{E}_N(x,t)\vert^p
{\bm 1}_{\vert\mathcal{E}_N(x,t)\vert\le f_{p-1}(N)}\, d^dx\le f_{p-1}(N)^p
\ll\frac{1}{\vert\Lambda\vert}\int_{\Lambda}\vert\mathcal{E}_N(x,t)\vert^p\, d^dx,
$$
hence
\begin{equation}\label{eq1.4}
\frac{1}{\vert\Lambda\vert}\int_{\Lambda}\vert\mathcal{E}_N(x,t)\vert^p\, d^dx\sim
\frac{1}{\vert\Lambda\vert}\int_{\Lambda}\vert\mathcal{E}_N(x,t)\vert^p
{\bm 1}_{\vert\mathcal{E}_N(x,t)\vert > f_{p-1}(N)}\, d^dx,
\end{equation}
almost surely as $N\rightarrow +\infty$. Equations\ (\ref{eq1.4}) means that when $N\rightarrow +\infty$, the space average of $\vert\mathcal{E}_N(x,t)\vert^p$ is almost surely determined by the region of $\Lambda$ in which $\vert\mathcal{E}_N(x,t)\vert> f_{p-1}(N)$. Now, we must prove that this region gets smaller and smaller as $N\rightarrow +\infty$. This is easily done for $p\ge 2$ as it follows immediately from\ (\ref{eq1.3b}) that
\begin{equation}\label{eq1.5}
\frac{1}{\vert\Lambda\vert}
\int_{\Lambda}{\bm 1}_{\vert\mathcal{E}_N(x,t)\vert >f_{p-1}(N)}d^dx\le
\frac{1}{f_{p-1}(N)^{p-1}\vert\Lambda\vert}
\int_{\Lambda}\vert\mathcal{E}_N(x,t)\vert^{p-1} d^dx \rightarrow 0,
\end{equation}
almost surely as $N\rightarrow +\infty$. For $p=1$ assume that there exists $\varepsilon >0$ small enough such that $\limsup_{N\rightarrow +\infty}\vert\Lambda\vert^{-1}\int_{\Lambda}\vert\mathcal{E}_N(x,t)\vert^{\varepsilon} d^dx <+\infty$ with probability one, (this will be the case for the type of $\mathcal{E}_N$ considered in this paper). Then,
\begin{equation}\label{eq1.6}
\frac{1}{\vert\Lambda\vert}
\int_{\Lambda}{\bm 1}_{\vert\mathcal{E}_N(x,t)\vert >f_0(N)}d^dx\le
\frac{1}{f_0(N)^{\varepsilon}\vert\Lambda\vert}
\int_{\Lambda}\vert\mathcal{E}_N(x,t)\vert^{\varepsilon} d^dx\rightarrow 0,
\end{equation}
almost surely as $N\rightarrow +\infty$, which completes the proof that $\mathcal{E}_N(x,t)$ is intermittent when\ (\ref{eq1.2}) holds.

In this paper we determine the intermittency threshold defined as the smallest value of $\lambda$ above which $\mathcal{E}_N(x,t)$ is intermittent, assuming ergodicity as $N\rightarrow +\infty$ for $\lambda$ small enough. By ergodicity as $N\rightarrow +\infty$ we mean that if $\lim_{N\rightarrow +\infty}\langle\vert\mathcal{E}_N\vert\rangle$ exists, then
\begin{equation}\label{eq1.7}
\lim_{N\rightarrow +\infty}\frac{1}{\vert\Lambda\vert}
\int_{\Lambda}\vert\mathcal{E}_N(x,t)\vert\, d^dx=
\lim_{N\rightarrow +\infty}\langle\vert\mathcal{E}_N\vert\rangle ,
\end{equation}
almost surely, where $\langle\cdot\rangle$ denotes average over the realizations of $S_N$. We consider the simplest case $\vert m\vert =+\infty$, i.e.\ (\ref{eq1.1}) without the $\Delta\mathcal{E}_N$ term, for a time-independent driver $S_N(x,t)\equiv S_N(x)$.

The outline of the paper is as follows. In Section\ \ref{sec2} we specify our model. Section\ \ref{sec3} deals with the asymptotic behavior of $\langle\mathcal{E}_N(x,t)\rangle$ for large $N$. Intermittency of $\mathcal{E}_N(x,t)$ and loss of ergodicity are investigated in Section\ \ref{sec4}.
%
%
\section{Model and definitions}\label{sec2}
We assume that $\Lambda$ is a $d$-dimensional torus of length $L$ and volume $\vert\Lambda\vert =L^d$. In the following we take $L=1$ without loss of generality. For any given $N\in {\mathbb N}$, let $\mathcal{A}_N=\left\lbrace n\in\mathbb{Z}^d : n\in\lbrack -N,N\rbrack^d \right\rbrace$, ${\rm Card}\, \mathcal{A}_N =(2N+1)^d$, and assume that $S_N$ is of the form
\begin{equation}\label{eq2.1}
S_N(x)=\frac{1}{(2N+1)^{d/2}}\sum_{n\in\mathcal{A}_N}
\exp\left\lbrack i(\theta_n+2\pi n\cdot x)\right\rbrack ,
\end{equation}
where the $\theta_n$ are i.i.d. random phases uniformly distributed over $\lbrack 0,2\pi\lbrack$. In the terminology of the theory of random series of functions, $S_N$ is called a Steinhaus (finite) series\ \cite{Kah}. The average over the realizations of the $\theta_n$ is denoted by $\langle\cdot\rangle_\theta$, or simply by $\langle\cdot\rangle$ if there is no risk of confusion.

In the limit $\vert m\vert =+\infty$ and for $S_N$ given by\ (\ref{eq2.1}), the solution to\ (\ref{eq1.1}) reduces to $\mathcal{E}_N(x,t)=\exp(\lambda t\vert S_N(x)\vert^2)$. Introducing the average gain factor $g\equiv\lambda t\langle\vert S_N(x)\vert^2\rangle =\lambda t$ and using the fact that $\mathcal{E}_N(x,t)$ is actually a function of $x$ and $g$ only, one is led to study the intermittency of the field
\begin{equation}\label{eq2.2}
\mathcal{E}_N(x,g)=\exp\left\lbrack g\vert S_N(x)\vert^2\right\rbrack,
\end{equation}
as $N\rightarrow +\infty$. The onset of intermittency will be characterized by the intermittency threshold, $g_{th}$, defined by
\begin{equation}\label{eq2.3}
g_{th}=\inf\lbrace g>0:{\mathcal E}_N(x,g){\rm\ is\ intermittent}\rbrace .
\end{equation}
%
%
\section{Asymptotic behavior of $\bm{\langle\mathcal{E}_N(x,g)\rangle}$ for large $\bm{N}$}\label{sec3}
As we will see in the following, the intermittency properties of $\mathcal{E}_N(x,g)$ depend on the behavior of $\langle\mathcal{E}_N(x,g)\rangle$ for large $N$. This behavior is summarized in the following two lemmas.
\begin{lemma}\label{lem1}
If $g<1$, then $\forall\, N\ge 0$, $\langle\mathcal{E}_N(x,g)\rangle\le (1-g)^{-1}$, and
\begin{equation}\label{eq3.5}
\lim_{N\rightarrow +\infty}\langle\mathcal{E}_N(x,g)\rangle =\frac{1}{1-g}.
\end{equation}
\end{lemma}
{\it Proof.}
Let $h$ be a complex-valued zero-mean Gaussian random variable with $\langle h^2\rangle =0$ and $\langle\vert h\vert^2\rangle =(2N+1)^{-d}$. Write $\exp\left\lbrack g\vert S_N(x)\vert^2\right\rbrack$ as
\begin{equation}\label{eq3.0}
\exp\left\lbrack g\vert S_N(x)\vert^2\right\rbrack =
\left\langle {\rm e}^{\sqrt{g}(2N+1)^{d/2}\, \lbrack S_N(x)h^\ast +S_N(x)^\ast h\rbrack}\right\rangle_h.
\end{equation}
Let $u=\vert h\vert^2$ and define
\begin{equation}\label{eq3.1}
f_g(u)=u-\ln\, {\rm I_0}(2\sqrt{gu}),
\end{equation}
where ${\rm I_0}$ is the modified Bessel function of zero order. From\ (\ref{eq2.1}) and the integral representation of ${\rm I_0}$\ \cite{AS}, one gets,
\begin{eqnarray}\label{eq3.2}
\langle\mathcal{E}_N(x,g)\rangle &=&
\left\langle\left\langle {\rm e}^{\sqrt{g}(2N+1)^{d/2}\, \lbrack S_N(x)h^\ast +S_N(x)^\ast h\rbrack}\right\rangle_h\right\rangle_\theta \nonumber\\
&=&\left\langle\left\langle {\rm e}^{\sqrt{g}(2N+1)^{d/2}\, \lbrack S_N(x)h^\ast +S_N(x)^\ast h\rbrack}\right\rangle_\theta\right\rangle_h \nonumber\\
&=&\frac{(2N+1)^d}{\pi}\int_{-\infty}^{+\infty}\int_{-\infty}^{+\infty}
{\rm e}^{-(2N+1)^df_g(\vert h\vert^2)}dh_rdh_i \nonumber\\
&=&(2N+1)^d\int_0^{+\infty}{\rm e}^{-(2N+1)^d f_g(u)}du.
\end{eqnarray}
If $g<1$, it follows from\ (\ref{eq3.2}) and the inequality $\ln\, {\rm I_0}(2\sqrt{gu})\le gu$ that $\langle\mathcal{E}_N(x,g)\rangle$ is bounded above by
\begin{equation}\label{eq3.3}
\langle\mathcal{E}_N(x,g)\rangle\le
(2N+1)^d\int_0^{+\infty}{\rm e}^{-(2N+1)^d (1-g)u}du=\frac{1}{1-g}.
\end{equation}
Furthermore, it can be easily checked that $f_g(u)$ is minimum at the boundary $u=0$ with $f_g(0)=0$ and $f_g^\prime(0)=1-g>0$. The asymptotic behavior of\ (\ref{eq3.2}) in the large $N$ limit is thus determined by the vicinity of $u=0$ and one finds
\begin{equation}\label{eq3.4}
\langle\mathcal{E}_N(x,g)\rangle=
\frac{1}{1-g}\left\lbrack 1-O\left(\frac{1}{N^d}\right)\right\rbrack\ \ \ \ (N\rightarrow +\infty),
\end{equation}
hence\ (\ref{eq3.5}), which completes the proof of Lemma\ \ref{lem1}.\ $\square$
\begin{lemma}\label{lem2}
If $g>1$, then $\exists\, \gamma_g >0$ such that $\langle\mathcal{E}_N(x,g)\rangle$ behaves like $(2N+1)^{d/2}\exp\lbrack\gamma_g (2N+1)^d\rbrack$ as $N\rightarrow +\infty$.
\end{lemma}
{\it Proof.}
It can be checked from\ (\ref{eq3.1}) that if $g>1$ there exists a unique number $u_0>0$ such that $f_g(u)$ reaches its minimum at $u=u_0$, with $f_g(u_0)<0$, $f_g^\prime(u_0)=0$, and $f_g^{\prime\prime}(u_0)>0$. Write $\gamma_g =-f_g(u_0)>0$. The asymptotic behavior of\ (\ref{eq3.2}) in the large $N$ limit is now determined by the vicinity of $u=u_0$, yielding
\begin{equation}\label{eq3.6}
\langle\mathcal{E}_N(x,g)\rangle\sim
\sqrt{\frac{2\pi}{f^{\prime\prime}_g(u_0)}}(2N+1)^{d/2}
\left\lbrack 1+O\left(\frac{1}{N^{d/2}}\right)\right\rbrack
\exp\lbrack\gamma_g (2N+1)^d\rbrack\ \ \ \ (N\rightarrow +\infty),
\end{equation}
which completes the proof of Lemma\ \ref{lem2}.\ $\square$

\bigskip
It follows from\ (\ref{eq3.5}) and\ (\ref{eq3.6}) that there is a transition from a regime where $\lim_{N\rightarrow +\infty}\langle\mathcal{E}_N(x,g)\rangle <+\infty$ to a regime where $\lim_{N\rightarrow +\infty}\langle\mathcal{E}_N(x,g)\rangle =+\infty$ as $g$ increases past $g=1$. The value $g=1$ (or, more exactly, $\lambda =1/t$) is the counterpart of what we called the ``critical coupling" in\ \cite{MCL}.

Note also that $\langle\mathcal{E}_N(x,g)\rangle$ does not depend on $x$. In the following we will write $\langle\mathcal{E}_N(g)\rangle$ for $\langle\mathcal{E}_N(x,g)\rangle$.
%
%
\section{Intermittency of $\bm{\mathcal{E}_N(x,g)}$ and loss of ergodicity}\label{sec4}
In this section we investigate the intermittency of $\mathcal{E}_N(x,g)$ as $N\rightarrow +\infty$, assuming ergodicity for $g<1$ where $\lim_{N\rightarrow +\infty}\langle\mathcal{E}_N(g)\rangle$ exists.
\begin{proposition}\label{prop1}
If $\forall\ g<1$, one has
\begin{equation}\label{eq5.1}
\lim_{N\rightarrow +\infty}\int_{\Lambda}\mathcal{E}_N(x,g)\, d^dx=
\lim_{N\rightarrow +\infty}\langle\mathcal{E}_N(g)\rangle =
\frac{1}{1-g},
\end{equation}
almost surely, then $\forall\ g>1$,
\begin{equation}\label{eq5.2}
\lim_{N\rightarrow +\infty}\int_{\Lambda}\mathcal{E}_N(x,g)\, d^dx=
+\infty ,
\end{equation}
and
\begin{equation}\label{eq5.3}
\lim_{N\rightarrow +\infty}\frac{1}{\langle\mathcal{E}_N(g)\rangle}\int_{\Lambda}\mathcal{E}_N(x,g)\, d^dx=0 ,
\end{equation}
almost surely.
\end{proposition}
[For $\vert\Lambda\vert\ne 1$, the left-hand side of Eqs.\ (\ref{eq5.1}) to\ (\ref{eq5.3}) is divided by $\vert\Lambda\vert$]. Assuming ergodicity for $g<1$ [Eqs.(\ref{eq5.1})], one finds that $\int_{\Lambda}\mathcal{E}_N(x,g)\, d^dx$ is not asymptotic to $\langle\mathcal{E}_N(g)\rangle$ as $N\rightarrow +\infty$ if $g>1$ [Eq.(\ref{eq5.3})]. In this asymptotic sense one can say that ergodicity is lost as $g$ increases past the value $g=1$. Note also that\ (\ref{eq5.1}) ensures that the assumption we made to prove\ (\ref{eq1.6}) is fulfilled: use $\mathcal{E}_N(x,g)^\varepsilon =\mathcal{E}_N(x,g\varepsilon)$ and take $\varepsilon <1/g$.

\bigskip
\noindent {\it Proof.}
From\ (\ref{eq2.2}) it follows that for every $N\ge 0$ and every realization of the $\theta_n$, $\int_{\Lambda}\mathcal{E}_N(x,g)\, d^dx$ is a non decreasing function of $g$. Thus, for every $g>1$ and $\varepsilon>0$,
$$
\int_{\Lambda}\mathcal{E}_N(x,g)\, d^dx\ge\int_{\Lambda}\mathcal{E}_N(x,1-\varepsilon)\, d^dx,
$$
and by Eq.\ (\ref{eq5.1}),
\begin{equation}\label{eq5.4}
\liminf_{N\rightarrow +\infty}\int_{\Lambda}\mathcal{E}_N(x,g)\, d^dx
\ge\lim_{N\rightarrow +\infty}\int_{\Lambda}\mathcal{E}_N(x,1-\varepsilon)\, d^dx
=\frac{1}{\varepsilon},
\end{equation}
almost surely. Letting $\varepsilon\rightarrow 0$ yields\ (\ref{eq5.2}).

We now prove the limit\ (\ref{eq5.3}). From the control\ (\ref{eqa.6}) with $\alpha =3/4$ it follows that for $N$ large enough,
\begin{eqnarray}\label{eq5.5}
\int_{\Lambda}\mathcal{E}_N(x,g)\, d^dx &\le&
\exp\left\lbrack g\sup_{x\in\Lambda}\vert S_N(x)\vert^2\right\rbrack \nonumber \\
&<&\exp\left\lbrack 2g(2N+1)^{d/2}\right\rbrack ,
\end{eqnarray}
almost surely, and by Lemma\ \ref{lem2},
\begin{equation}\label{eq5.6}
\lim_{N\rightarrow +\infty}\frac{1}{\langle\mathcal{E}_N(g)\rangle}
\int_{\Lambda}\mathcal{E}_N(x,g)\, d^dx=0,
\end{equation}
with probability one, which completes the proof of Proposition\ \ref{prop1}. $\square$
\begin{proposition}[transition to intermittency]\label{prop2}
Under the same ergodicity assumption as in Proposition\ \ref{prop1}, $g_{th}=1$.
\end{proposition}
{\it Proof.}
First, we prove $g_{th}\ge 1$. If $\mathcal{E}_N(x,g)$ is intermittent, then for every integer $p\ge 1$ and almost all the realizations of $\mathcal{E}_N(x,g)$, there exists $f_{p-1}(N)$, with $f_{p-1}(N)\rightarrow +\infty$ as $N\rightarrow +\infty$, such that
\begin{equation}\label{eq5.7}
\int_{\Lambda}\mathcal{E}_N(x,g)^p\, d^dx\sim
\int_{\Lambda}\mathcal{E}_N(x,g)^p
{\bm 1}_{\vert\mathcal{E}_N(x,g)\vert > f_{p-1}(N)}\, d^dx,
\end{equation}
as $N\rightarrow +\infty$. From H\"older's inequality and
$$
\int_{\Lambda}{\bm 1}_{\mathcal{E}_N(x,g)>f_0(N)}d^dx
\le\frac{1}{f_0(N)}\int_{\Lambda}\mathcal{E}_N(x,g)\, d^dx,
$$
one gets, $\forall\varepsilon >0$,
\begin{eqnarray}\label{eq5.8}
&&\int_{\Lambda}\mathcal{E}_N(x,g)
{\bm 1}_{\mathcal{E}_N(x,g)>f_0(N)}\, d^dx \nonumber \\
&&\le\left\lbrack\int_{\Lambda}
\mathcal{E}_N(x,g)^{1+\varepsilon} d^dx\right\rbrack^{\frac{1}{1+\varepsilon}}
\left\lbrack\int_{\Lambda}
{\bm 1}_{\mathcal{E}_N(x,g)>f_0(N)}\, d^dx\right\rbrack^{\frac{\varepsilon}{1+\varepsilon}} \nonumber\\
&&=\left\lbrack\int_{\Lambda}
\mathcal{E}_N(x,g(1+\varepsilon))\, d^dx\right\rbrack^{\frac{1}{1+\varepsilon}}
\left\lbrack\int_{\Lambda}
{\bm 1}_{\mathcal{E}_N(x,g)>f_0(N)}\, d^dx\right\rbrack^{\frac{\varepsilon}{1+\varepsilon}} \\
&&\le\frac{1}{f_0(N)^{\frac{\varepsilon}{1+\varepsilon}}}
\left\lbrack\int_{\Lambda}
\mathcal{E}_N(x,g(1+\varepsilon))\, d^dx\right\rbrack^{\frac{1}{1+\varepsilon}}
\left\lbrack\int_{\Lambda}
\mathcal{E}_N(x,g)\, d^dx\right\rbrack^{\frac{\varepsilon}{1+\varepsilon}}. \nonumber 
\end{eqnarray}
Take $g<1/(1+\varepsilon)$. By\ (\ref{eq5.1}), the two brackets on the right-hand side of\ (\ref{eq5.8}) are almost surely bounded and $\int_{\Lambda}\mathcal{E}_N(x,g){\bm 1}_{\mathcal{E}_N(x,g) > f_0(N)}\, d^dx\rightarrow 0$ almost surely as $N\rightarrow +\infty$. This is in contradiction with\ (\ref{eq5.7}) for $p=1$. Thus, for every $g<1/(1+\varepsilon)$, $\mathcal{E}_N(x,g)$ is not intermittent and by taking $\varepsilon >0$ arbitrarily small one obtains $g_{th}\ge 1$.

We now prove  $g_{th}\le 1$. To this end we prove that\ (\ref{eq1.2}) is fulfilled if $g>1$. Using H\"older's and Jensen's inequalities successively, one finds that for every integer $p\ge 1$ and $\forall\, 0<\varepsilon <1$,
\begin{eqnarray}\label{eq5.9}
&&\left\lbrack
\int_{\Lambda}\mathcal{E}_N(x,g)^pd^dx
\right\rbrack^{\frac{1}{p}}=
\left\lbrack
\int_{\Lambda}\mathcal{E}_N(x,g)^{p(1-\varepsilon)}
\mathcal{E}_N(x,g)^{p\varepsilon} d^dx
\right\rbrack^{\frac{1}{p}} \nonumber\\
&&\le\left\lbrack
\int_{\Lambda}\mathcal{E}_N(x,g)^{(p+1)(1-\varepsilon)}d^dx
\right\rbrack^{\frac{1}{p+1}}
\left\lbrack
\int_{\Lambda}\mathcal{E}_N(x,g)^{p(p+1)\varepsilon} d^dx
\right\rbrack^{\frac{1}{p(p+1)}} \nonumber\\
&&=\left\lbrack
\int_{\Lambda}\mathcal{E}_N(x,g)^{(p+1)(1-\varepsilon)}d^dx
\right\rbrack^{\frac{1}{p+1}}
\left\lbrack
\int_{\Lambda}\mathcal{E}_N(x,gp(p+1)\varepsilon)\, d^dx
\right\rbrack^{\frac{1}{p(p+1)}} \\
&&\le\left\lbrack
\int_{\Lambda}\mathcal{E}_N(x,g)^{p+1}d^dx
\right\rbrack^{\frac{1-\varepsilon}{p+1}}
\left\lbrack
\int_{\Lambda}\mathcal{E}_N(x,gp(p+1)\varepsilon)\, d^dx
\right\rbrack^{\frac{1}{p(p+1)}}, \nonumber
\end{eqnarray}
which gives,
\begin{eqnarray}\label{eq5.10}
&&\left\lbrack
\int_{\Lambda}\mathcal{E}_N(x,g)^{p+1}d^dx
\right\rbrack^{\frac{1}{p+1}}
\left\lbrack
\int_{\Lambda}\mathcal{E}_N(x,g)^pd^dx
\right\rbrack^{-\frac{1}{p}} \nonumber\\
&&\ge\left\lbrack
\int_{\Lambda}\mathcal{E}_N(x,g)^{p+1}d^dx
\right\rbrack^{\frac{\varepsilon}{p+1}}
\left\lbrack
\int_{\Lambda}\mathcal{E}_N(x,gp(p+1)\varepsilon)\, d^dx
\right\rbrack^{-\frac{1}{p(p+1)}} \\
&&=\left\lbrack
\int_{\Lambda}\mathcal{E}_N(x,g(p+1))\, d^dx
\right\rbrack^{\frac{\varepsilon}{p+1}}
\left\lbrack
\int_{\Lambda}\mathcal{E}_N(x,gp(p+1)\varepsilon)\, d^dx
\right\rbrack^{-\frac{1}{p(p+1)}}. \nonumber
\end{eqnarray}
For every $g>1$, take $0<\varepsilon <\lbrack gp(p+1)\rbrack^{-1}$. By\ (\ref{eq5.1}) and\ (\ref{eq5.2}), one has, with probability one,
\begin{equation}\label{eq5.11}
\lim_{N\rightarrow +\infty}
\int_{\Lambda}\mathcal{E}_N(x,gp(p+1)\varepsilon)\, d^dx <+\infty ,
\end{equation}
and
\begin{equation}\label{eq5.12}
\lim_{N\rightarrow +\infty}
\int_{\Lambda}\mathcal{E}_N(x,g(p+1))\, d^dx =+\infty .
\end{equation}
Injecting\ (\ref{eq5.11}) and\ (\ref{eq5.12}) into the right-hand side of\ (\ref{eq5.10}), one gets
\begin{equation}\label{eq5.13}
\left\lbrack
\int_{\Lambda}\mathcal{E}_N(x,g)^pd^dx
\right\rbrack^{\frac{1}{p}}
\ll
\left\lbrack
\int_{\Lambda}\mathcal{E}_N(x,g)^{p+1}d^dx
\right\rbrack^{\frac{1}{p+1}},
\end{equation}
almost surely. It remains to prove the first inequality\ (\ref{eq1.2}), which is immediate by\ (\ref{eq5.2}). Thus,\ (\ref{eq1.2}) is fulfilled and $\mathcal{E}_N(x,g)$ is intermittent for every $g>1$. This implies $g_{th}\le 1$, which completes the proof of Proposition\ \ref{prop2}. $\square$
%
%
\section{Summary and perspectives}\label{sec5}
We have investigated intermittency of $\mathcal{E}_N(x,g)=\exp\lbrack g\vert S_N(x)\vert^2\rbrack$ as $N\rightarrow +\infty$, when $S_N(x)$ is given by the Steinhaus series\ (\ref{eq2.1}). Assuming ergodicity of $\mathcal{E}_N(x,g)$ as $N\rightarrow +\infty$ for $g<1$, where $\lim_{N\rightarrow +\infty}\langle\mathcal{E}_N(g)\rangle$ exists, we have proved the existence of a transition to intermittency as $g$ increases past the threshold $g_{th}=1$ (Proposition\ \ref{prop2}). This transition goes together with a loss of ergodicity in the sense of a transition from (assumed) ergodicity at $g<g_{th}$ to  a regime where $\lim_{N\rightarrow +\infty}\lbrack\vert\Lambda\vert\langle\mathcal{E}_N(g)\rangle\rbrack^{-1}\int_{\Lambda}\mathcal{E}_N(x,g)\, d^dx=0$ at $g>g_{th}$ (Propositions\ \ref{prop1}). Proving ergodicity of $\mathcal{E}_N(x,g)$ as $N\rightarrow +\infty$ for $g<1$ is another problem that we are unable to solve at the present time.

The next step toward a study of the solution to\ (\ref{eq1.1}) will consist in allowing for a time dependent $S_N$ in the simpler Laplacian free case ($\vert m\vert =+\infty$). What is to be expected in this setting can be conjectured in view of the results obtained in this paper. Indeed, it is easily seen that the intermittency threshold of Proposition\ \ref{prop2} corresponds to the critical coupling defined as the smallest $g$ at which $\langle\mathcal{E}_N(x,g)\rangle$ would diverge if $S_N(x)$ was a Gaussian r.v. with $\langle S_N(x)\rangle =\langle S_N(x)^2\rangle =0$ and $\langle\vert S_N(x)\vert^2\rangle =1$. Intuitively, this could have been expected from the CLT according to which, for any fixed $x\in\mathbb{R}^d$,\ (\ref{eq2.1}) tends in law to such a Gaussian r.v. as $N\rightarrow +\infty$. Now, if $\forall\, x\in\mathbb{R}^d$, $S_N(x,t)$ tends in law to a Gaussian random function of $t$ as $N\rightarrow +\infty$, it is not unreasonable to conjecture that the intermittency threshold for $\mathcal{E}_N(x,t)$ should be given by the critical coupling in this case too. Namely, for any given $t>0$,
$$
\lambda_{th}(t)=\frac{1}{\mu_1(t)},
$$
where $\mu_1(t)$ is the largest eigenvalue of the covariance of $S_N(x,\tau)$ for $0\le\tau\le t$. Proving this conjecture goes through the resolution of specific technical problems inherent in $\mathcal{E}_N(x,t)$ being now a functional of $S_N(x,t)$. This will be the subject of a forthcoming paper.

We will then be ready to tackle the study of intermittency of the solution to\ (\ref{eq1.1}) with a finite $m$. This more difficult problem will presumably require the use of the distributional formulation of\ \cite{MCL}.
%
%
\appendix*
\section{Controlling the excursion of $\bm{\vert S_N(x)\vert}$}
Pave $\Lambda$ with $d$-dimensional cubes, $\Lambda_i$, of length $\ell\le 1$, with $\ell^{-1}\in\mathbb{N}$ and $1\le i\le\ell^{-d}$. Let $x_i\in\Lambda$ denote the center of $\Lambda_i$. For every $x\in\Lambda_i$ one gets
\begin{eqnarray*}
\left\vert\vert S_N(x)\vert^2 -\vert S_N(x_i)\vert^2\right\vert
&=&\left\vert\sum_{n,m\in\mathcal{A}_N}
\frac{{\rm e}^{i(\theta_n-\theta_m)}}{(2N+1)^d}
\left({\rm e}^{2i\pi (n-m)\cdot x}
-{\rm e}^{2i\pi (n-m)\cdot x_i}\right)\right\vert \nonumber \\
&\le&\frac{2}{(2N+1)^d}\sum_{n,m\in\mathcal{A}_N}\left\vert\sin\left\lbrack
\pi (n-m)\cdot (x-x_i)\right\rbrack\right\vert \\
&\le&\frac{2\pi}{(2N+1)^d}\sum_{n,m\in\mathcal{A}_N}
\vert n-m\vert\vert x-x_i\vert . \nonumber
\end{eqnarray*}
Since $\vert n-m\vert\le 2N\sqrt{d}$ and $\vert x-x_i\vert\le\ell\sqrt{d}$ one has $\forall\, x\in\Lambda_i$,
\begin{equation}\label{eqa.1}
\left\vert\vert S_N(x)\vert^2 -\vert S_N(x_i)\vert^2\right\vert\le
4\pi d\ell N(2N+1)^d .
\end{equation}
Fix $1/2<\alpha\le 1$ and take
\begin{equation}\label{eqa.2}
\ell^{-1}={\rm int}(4\pi d+1)N(2N+1)^{2d(1-\alpha)}.
\end{equation}
Then,\ (\ref{eqa.1}) is bounded by
\begin{equation}\label{eqa.3}
\left\vert\vert S_N(x)\vert^2 -\vert S_N(x_i)\vert^2\right\vert\le (2N+1)^{d(2\alpha -1)}.
\end{equation}
Let $z\in\mathbb{C}$ and write $t\equiv\vert z\vert$ and $\varphi\equiv {\rm Arg}(z)-{\rm Arg}(S_N(x))$. For every $x\in\Lambda$, one has
$$
\left\langle\exp\lbrack zS_N(x)^\ast +c\cdot c\cdot\rbrack\right\rangle =
\exp\left\lbrack (2N+1)^d\ln\, {\rm I_0}\left(\frac{2t}{(2N+1)^{d/2}}\right)\right\rbrack ,
$$
and
\begin{eqnarray*}
\left\langle\exp\lbrack zS_N(x)^\ast +c\cdot c\cdot\rbrack\right\rangle &=&
\left\langle\exp\lbrack 2t\vert S_N(x)\vert\cos\varphi\rbrack\right\rangle \\
&\ge&\left\langle\exp\lbrack 2t\vert S_N(x)\vert\cos\varphi\rbrack
\bm{1}_{-\pi /3\le\varphi\le\pi /3}\bm{1}_{\vert S_N(x)\vert\ge (2N+1)^{d(\alpha -1/2)}}
\right\rangle \\
&\ge&\frac{1}{3}\exp\left\lbrack t(2N+1)^{d(\alpha -1/2)}\right\rbrack
P\left(\vert S_N(x)\vert\ge (2N+1)^{d(\alpha -1/2)}\right).
\end{eqnarray*}
Thus,
\begin{eqnarray*}
&&P\left(\vert S_N(x)\vert\ge (2N+1)^{d(\alpha -1/2)}\right) \\
&&\le
3\exp\left\lbrack (2N+1)^d\ln\, {\rm I_0}\left(\frac{2t}{(2N+1)^{d/2}}\right)
-t(2N+1)^{d(\alpha -1/2)}\right\rbrack \\
&&\le 3\exp\left\lbrack t^2-t(2N+1)^{d(\alpha -1/2)}\right\rbrack ,
\end{eqnarray*}
where we have used the inequality $\ln\, {\rm I_0}(2s)\le s^2$, and by taking $t=(2N+1)^{d(\alpha -1/2)}/2$ one gets
\begin{equation}\label{eqa.4}
P\left(\vert S_N(x)\vert\ge (2N+1)^{d(\alpha -1/2)}\right)\le
3\exp\left\lbrack -\frac{1}{4}(2N+1)^{d(2\alpha -1)}\right\rbrack .
\end{equation}
From\ (\ref{eqa.3}), it follows that if $\vert S_N(x_i)\vert <(2N+1)^{d(\alpha -1/2)}$ for every $i\le \ell^{-d}$, then  $\vert S_N(x)\vert <\sqrt{2}(2N+1)^{d(\alpha -1/2)}$ for every $x\in\Lambda$. Therefore, using\ (\ref{eqa.4}),
\begin{eqnarray}\label{eqa.5}
P\left(\sup_{x\in\Lambda}\vert S_N(x)\vert\ge \sqrt{2}(2N+1)^{d(\alpha -1/2)}\right)&\le&
P\left(\bigcup_{i=1}^{\ell^{-d}}\lbrace\vert S_N(x_i)\vert\ge (2N+1)^{d(\alpha -1/2)}\rbrace\right) \nonumber\\
&\le& 3\ell^{-d}\exp\left\lbrack -\frac{1}{4}(2N+1)^{d(2\alpha -1)}\right\rbrack .
\end{eqnarray}
Now, by\ (\ref{eqa.2}) $\ell^{-d}$ diverges algebraically in $N$ as $N\rightarrow +\infty$, and since $\alpha >1/2$ the right-hand side of the last inequality\ (\ref{eqa.5}) tends to zero faster than any power of $N$ as $N\rightarrow +\infty$. Thus,
$$
\sum_{N=1}^{+\infty}
P\left(\sup_{x\in\Lambda}\vert S_N(x)\vert\ge \sqrt{2}(2N+1)^{d(\alpha -1/2)}\right)<+\infty ,
$$
and by the Borel-Cantelli lemma,
\begin{equation}\label{eqa.6}
P\left(\limsup_{N\rightarrow +\infty}\sup_{x\in\Lambda}\vert S_N(x)\vert <\sqrt{2}(2N+1)^{d(\alpha -1/2)}\right)=1.
\end{equation}
%
%
%

%
%
\end{document}